# Recognition of Blockchain-based Multisignature E-Awards

A.J. Santos[1]

**Abstract**: With blockchain technology, information is recorded in a permanent distributed ledger that is maintained by multiple computers in a peer-to-peer network. There is no central authority that can alter records or change network consensus rules. Such technology could be utilized for voting, title transfers, issuance of company shares, document notarization, but currently, the most popular use-case are virtual currencies. An interesting feature that some virtual currencies have is a multisignature ("multisig") protocol that requires the electronic signatures from more than one private key to initiate a transfer of funds. Raw data of a multisig transaction may be recognized as an arbitral award under the New York Convention, where the law of England is the *lex arbitri* and parties have opted-out of a reasoned award.

## I. INTRODUCTION

Multisig is a trust-less method for conducting electronic transactions that could possibly be used for international trade. For instance, a seller and buyer can set-up a multisig account and nominate a third party. If goods are delivered, funds are released from the multisig account with the signatures of the buyer and seller. In case of a dispute, the third party could adjudicate the dispute and sign the release of funds in favor of the prevailing party. However, the legal effect of the act of releasing funds from a multisig scheme has not been fully explored. It is of vital importance for resolved disputes to be final and binding. Businesses operate across many different jurisdictions and having an innovative method to finally resolve disputes would help decrease transactional costs and bolster certainty in an electronic trading environment. For this to occur, multisig needs to be re-imagined not as a simple transaction, but an arbitration process. The resulting multisig electronic award would have *res judicata* effect, where parties can rely on multisig to further the development of a case or to stop the reopening of a matter which has already been decided in a previous proceeding[2]. The *res judicata* effect of awards is an integral part of international commercial arbitration, which is evident in Article III of the New York Convention that requires courts in contracting states to recognize awards as binding[3].

This paper will determine whether a multisig transaction, arising out of an arbitration agreement, can be viewed as an electronic arbitral award that is capable of recognition under international arbitration law. In Part I, the legal framework for the form and content requirements of an arbitral award will be discussed. In Part II, we will touch on the law of electronic signatures and consider the authentication requirement of the New York Convention; and finally, in Part III, we will look at the possibility of characterizing multisig arbitration as an autonomous legal order. It is assumed the reader has a basic understanding of blockchain technology and asymmetric cryptography.

---

1   B.A. (UTSA), J.D. (STCL), Department of Private International Law, Ankara Yıldırım Beyazıt University, Faculty of Law, Contact: ajsantos@protonmail.com
2   Schaffstein, Silja (2016) The Doctrine of Res Judicata Before International Commercial Arbitral Tribunals, Oxford, Oxford University Press, pg. 210; Trans-Lex.org 'No. XIII.4.5 - Conclusive and preclusive effect of awards; res judicata' <https://www.trans-lex.org/970070> l.a.d. 04/18/2019.
3   *Id.*

## II. AWARD FORM AND CONTENT

An award can be defined as a record that evidences an arbitrator's decision, which is the consequence of evaluating opposing contentions between parties, weighing evidence and submissions[4]. The award puts an end to the arbitration process in whole, or in part, and finally settles issues, which cannot later be reopened or revised[5]. In making an award, arbitrators must act in accordance with the principles of natural justice[6], where each party is given a full opportunity to present their case and parties are treated equally[7]. The arbitrators must also use their best efforts to render an enforceable award[8]. Form and content requirements are governed by multiple sources that include: the arbitration agreement itself, national arbitration laws, and arbitration rules; furthermore, arbitrators commonly refer to non-binding guidelines on drafting awards published by a number of non-governmental organizations.

The concept of party autonomy is a key pillar of international arbitration law. Parties have the freedom of contract to resolve disputes privately under the terms chosen by them. The parties can customize the entire arbitration process in their image when drafting the arbitration agreement to the extent allowed by public policy considerations. Parties have the confidence that the arbitration process is "their" arbitration and that it will be conducted according to their will[9]. In this sense, an arbitrator's power to issue an award is founded upon the consent of the parties. As arbitration is a creature of consent[10], the arbitration agreement ought to be the primary source for ascertaining the form in which an award is given. Some national laws have taken this view. For instance, in the United Kingdom, Section 52(1) of the Arbitration Act 1996 provides that "parties are free to agree on the form of an award," similarly, Article 189(3)(1) of Switzerland's Federal Code on Private International Law indicates that parties can agree on award form[11].

As part of this freedom, parties could choose which arbitration rules are applicable. Some arbitration rules call for arbitral awards to be in a particular form. For instance, under ICC Arbitration Rules, Article 32(2), an award should state the reasons upon which it is based on. Additional information in awards are called for by guidelines like the ICC Award Checklist for Arbitrators, which require, *inter alia*, the quotation of the entire arbitration agreement, relevant choice-of-law clause, and a summary of procedural steps that haven been taken[12]. The Chartered Institute of Arbitrators[13] suggests an award should also contain the names and addresses of the arbitrators, a summary of the facts and procedure, and operative language[14]. While the International Bar Association recommends an award should contain the commercial registration number and nationality of corporations, the nationalities of the arbitrators, and a chronology of the events leading to the commencement of the dispute[15].

---

4   Turner, Ray (2005) Arbitration Awards: A Practical Approach, Hoboken, Wiley-Blackwell, pg. 3.
5   ICCA (2011) Guide to the Interpretation of the 1958 New York Convention: a Handbook for Judges, pg. 17.
6   Turner (2005), *supra* 4 at pg. 5.
7   UNCITRAL Model Law on International Commercial Arbitration (1985), with amendments as adopted in 2006, Article 18; Trans-Lex.org 'Principle XIII.3.1 - Arbitral due process' <https://www.trans-lex.org/969020> l.a.d. 04/18/2019.
8   Platte, Martin (2003) 'An Arbitrator's Duty to Render Enforceable Awards' Journal of International Arbitration, I:3, V:20, pp. 307-313; Trans-Lex.org 'Principle IV.6.5 - Best efforts undertakings' <https://www.trans-lex.org/932000> l.a.d. 04/18/2019.
9   Chatterjee, C. (2003) 'The Reality of The Party Autonomy Rule In International Arbitration', Journal of International Arbitration, V:20, I:6, pg. 540.
10  Moses, Margaret L. (2010) 'Arbitration Law: Who's In Charge', Seton Hall Law Review, V:40, I:1, pg. 154.
11  Original text in German: "Article 189(1) - Der Entscheid ergeht nach dem Verfahren und in der Form, welche die Parteien vereinbart haben."

There is a tendency for awards to be very detailed. This could be the case because arbitrators have a duty to ensure awards are enforceable[16], so will err on the side of caution and include more information than less in order to provide evidence that the parties were given a full opportunity to present their case and were treated equally[17]. These procedural safeguards are meant to ensure compliance with the fundamental principle of due process, a violation of which may lead to the setting aside of an award[18]. Considering this, we now turn to the question if an electronic award recorded on a blockchain could meet the form requirements. As an illustration, the author performed a multisig transaction on the Bitcoin testnet network, details found in Appendix A.

It is proposed that the raw data of a multisig transaction could represent a final and binding arbitral award providing that the parties select a seat of arbitration in a jurisdiction conducive to the will of the parties as it relates to the form of an award. The parties would need to agree in an arbitration agreement that either the United Kingdom or Switzerland is the seat of arbitration. Furthermore, the appointed arbitrators and parties would need to be named and associated with specific cryptocurrency wallet addresses in the agreement, since this would provide a means for identifying the parties and arbitrators in the resulting multisig transaction. With this information and the testimony of an expert witness, an enforcement court would have sufficient evidence to establish the existence of an award.

However, ICC Arbitration Rules and guidelines, as noted above, require the award to be reasoned and include other information besides the name of the parties and place of arbitration. There are limitations to the amount of data that can be included in a transaction. In Bitcoin, OP_RETURN opcode has a limit of 83 bytes of data for any given transaction[19]. The data size of a reasoned award can be in the hundreds of kilobytes range. Even if the data is broken up into multiple transactions, the burden of storing and verifying large amounts of data would be shifted to node operators and could ultimately adversely affect the economic sustainability of the blockchain in the long term. Moreover, a very verbose award on a public blockchain undermines confidentiality of the parties as anyone can look-up the information. A possible solution around this is to not upload the award, but rather run a SHA-256 checksum of an award that is saved as a single file and anchor the resulting hash to a transaction in order to timestamp the document on the blockchain, or alternatively, the award could be encrypted and stored off-chain into a distributed file system, *e.g.* IPFS, where the associated hash is added to a transaction in a similar fashion[20]. In both of these methods, the transaction does not represent the actual award, but is a reference marker to a document. In this case, a multisig transaction utilizing these methods would be incapable of being an arbitral award when a reasoned award is required.

---

12  International Chamber of Commerce, 'ICC Award Checklist' <https://iccwbo.org/publication/icc-award-checklist-1998-2012-icc-arbitration-rules> l.a.d. 04/17/2019.
13  The Chartered Institute of Arbitrators, 'Drafting Arbitral Awards Part I' <https://www.ciarb.org/media/4206/guideline-10-drafting-arbitral-awards-part-i-general-2016.pdf> l.a.d. 04/17/2019.
14  Operative language is a phrase that is indicative of a command, such as "we award," "we direct," "we order."
15  International Bar Association, 'Guidelines for Drafting International Arbitration Clauses' <https://www.ibanet.org/lpd/dispute_resolution_section/arbitration/projects.aspx> l.a.d. 04/17/2019.
16  *supra* note 8.
17  *supra* note 7.
18  New York Convention, Article V(b).
19  It should be noted that many members of the Bitcoin community object to the use of OP_RETURN to store arbitrary data. *See* Bistarelli, Stefano/Mercanti, Ivan/Santini, Francesco 'An Analysis of Non-standard Transactions', Front. Blockchain <https://doi.org/10.3389/fbloc.2019.00007> l.a.d. 08/13/2019.

On the other hand, there are some institutional rules that do allow parties to explicitly opt-out of a reasoned award[21] and given the concept of party autonomy, parties are able to prohibit arbitrators from following guidelines on the form of an award in the arbitration agreement. Another option is to adopt bespoke ad-hoc arbitration rules that would allow for electronic awards. In both approaches, most institutional rules, Article 31 of UNCITRAL Model Law, and Article 34 of UNCITRAL Arbitration Rules, call for awards to be "in writing" and "signed" by the arbitrators. One would have to consider the functional equivalence of paper documents in the electronic medium. This issue is particularly problematic for a party attempting to seek recognition under the New York Convention as Article IV(1)(a) requires a duly authenticated original award.

### III. ELECTRONIC SIGNATURES, WRITING REQUIREMENT, AND AUTHENTICATION

#### A. IS IT POSSIBLE FOR AN ARBITRATOR TO DIGITALLY SIGN AN ARBITRAL AWARD?

In the case of the European Union, the eIDAS Regulation[22] provides a legal framework for electronic signatures. The eIDAS Regulation was implemented in the United Kingdom[23] with the Electronic Identification and Trust Services for Electronic Transactions Regulations 2016/696, which superseded the Electronic Signatures Regulations 2002/318. The Electronic Communications Act 2000 ("ECA 2000") was amended with some provisions from the eIDAS Regulation. The ECA 2000 does not establish the inherent validity of electronic signatures and only confirms their admissibility[24]. According to Section 7(1)(a), in any legal proceedings, an electronic signature is admissible as evidence. Article 25 of the eIDAS Regulation, which gives electronic signatures legal effect[25], was not transpose to the ECA 2000. The drafters of the ECA 2000 might have wanted to give courts greater latitude in determining the proper use and the evidential value of electronic signatures[26]. Electronic signatures are defined in Section 7(2) as anything in electronic form "incorporated into or otherwise logically associated with any electronic communication or electronic data" and that "purports to be used by [an] individual creating it to sign." The second element thus indicates that there must be an intention to sign.

---

20 The author illustrates this method with the use of Monero, a privacy-centric cryptocurrency, and IPFS. *See* <https://github.com/monero-ecosystem/monero-pgp-messenger> l.a.d. 08/13/2019.
21 *See* Stockholm Chamber of Commerce Rules, Article 42. Hong Kong International Arbitration Centre Rules, Article 35. Although HKIAC Article 35.6 adds an additional form requirement of a seal affixed to the award.
22 Regulation (EU) No. 910/2014 of the European Parliament and of the Council of 23 July 2014 on Electronic Identification and Trust Services for Electronic Transactions in the Internal Market and Repealing Directive 1999/93/EC, 257 OJ L (2014) <http://data.europa.eu/eli/reg/2014/910/oj/eng> l.a.d. 04/17/2019.
23 Under European Union (Withdrawal) Act of 2018, most E.U. laws, including the eIDAS Regulation, would be retained and considered part of domestic law when U.K. exits the E.U.
24 A counter-view is expressed by Mr. Justice Popplewell in Bassano v Toft [2014] EWHC 377 (QB) ("[section 7] recognises the validity of such an electronic signature by providing that an electronic signature is admissible as evidence of authenticity").
25 eIDAS Regulation, Article 25(1), "an electronic signature shall not be denied legal effect and admissibility as evidence in legal proceedings solely on the grounds that it is in an electronic form or that it does not meet the requirements for qualified electronic signatures."
26 Law Commission, 'Report on Electronic Execution of Documents' para. 3.21-3.23, pg. 36.

An intention to sign was demonstrated in the case of *Goodman v. J. Eban Limited*[27], Mr. Goodman, a solicitor, used a rubber stamp to sign letters sent with his bill of costs. He kept the stamp locked in his room and was the only person who had access to the stamp. The Court of Appeal held that the use of such a mechanical signature is valid given that Mr. Goodman intended the rubber stamp to be regarded as a signature for signing letters. In quoting an older authority, *Bennett v. Brumfitt*, the court noted that a signature is not actually made by the hand alone, but with the use of some instrument. There is no distinction between using a pen, pencil, stamp or even a paint-brush, where the impression is made with the intent and purpose of signing a document. The court acknowledged that a stamp might not, on the face of it, carry the same assurance of authenticity as a signature written in the ordinary way with a pencil or pen, but stated that if there is doubt, a party could inquire whether the signer had personally signed the document.

Electronic signatures originating from a virtual currency wallet can be admitted as evidence and eventually found to be valid by a court. A Bitcoin address is a public key hash encoded as a base58 string[28]. When a sender creates a standard P2PKH[29] transaction containing instructions that would allow a recipient, whom controls the private key of the public key hash, to spend an output and includes a scriptSig[30] to authorize the release of funds, the sender is incorporating unique electronic data for the purpose of signing the transaction. When a sender broadcast the transaction to the network, they are manifesting an intent to hold the signatures incorporated into the data as valid. In considering the example in Appendix A, the identity of the wallet addresses can be linked to parties. If authenticity is at issue, a party can digitally sign a message with their wallet attesting to the possession and sole control over the wallet's private key[31]. To increase the legal certainty of the legal effect of electronic signatures, arbitrators may also e-mail parties a transaction ID ("TXID"), which is a reference number to a transaction on the blockchain, accompanied with a copy of "wet ink signatures" in pdf format.

## B. HOW COULD AN ELECTRONIC AWARD BE CONSIDERED TO BE "IN WRITING?"

Many jurisdictions have adopted the UNCITRAL Model Law on Electronic Commerce or similar laws[32], which view electronic records as valid and enforceable[33]. Where a law, or a legal obligation, requires information to be evidenced "in writing," this requirement can be met by an electronic record if the information contained therein is subsequently retrievable at a later time[34]. Furthermore, an electronic record could be considered to be "original" if: (a) there is reliable assurance of the integrity of the information from the time when it was first generated and (b) the information is capable of being represented[35].

---

27 Goodman v. J Eban LD, [1954] 1QB 550 (U.K.).
28 Bitcoin Project <https://bitcoin.org/en/transactions-guide#introduction> l.a.d. 04/17/2019.
29 Bitcoin Project <https://bitcoin.org/en/transactions-guide#term-p2pkh> l.a.d. 04/17/2019.
30 Bitcoin Project <https://bitcoin.org/en/glossary/signature-script> l.a.d. 04/17/2019.
31 Bitcoin Project <https://bitcoin.org/en/developer-reference#signmessage> l.a.d. 04/17/2019.
32 72 States and a total of 151 jurisdictions have adopted UNCITRAL Model Law on Electronic Commerce (1996) with additional article 5 bis as adopted in 1998 <https://uncitral.un.org/en/texts/ecommerce/modellaw/electronic_commerce/status> l.a.d. 04/17/2019. In the United States, the Electronic Signatures in Global and National Commerce Act (E-SIGN) follows a similar approach of the Model Law.
33 UNCITRAL Model Law on Electronic Commerce (1996), Article 5.
34 UNCITRAL Model Law on Electronic Commerce (1996), Article 6.
35 UNCITRAL Model Law on Electronic Commerce (1996), Article 8.

For the United Kingdom, under Section 7(c)(2) of the ECA 2000, an "electronic document" is defined as anything stored in electronic form, including text, and Section 7(c)(1), states that in any legal proceedings, an electronic document is admissible as evidence if it relates to the authenticity of an electronic transaction. Black's Law Dictionary defines a transaction as "[t]he act or an instance of conducting business or other dealings; esp., the formation, performance, or discharge of a contract[36]."

The definition of a transaction is broad enough to include raw data of a multisig transaction. The data results from the execution of a contractual obligation of multisig participants. Furthermore, the multisig transaction is made up of hexadecimal data with digits from 0 to 9 and characters A to F to represent 16 possible values[37]. If a single number or letter is converted to a different value, the TXID will be changed and the altered TXID would not be found in a blockchain explorer. The raw transaction data is a unique value that is validated as a true transaction by many computers in a network, in a sense, raw transaction data is more temper-resistant than traditional paper awards. The integrity of raw transaction data is reliably assured from the time when it is first committed to a block and the information is represented in a complete and unaltered state in a permanent distributed ledger. As such, a party attempting to prove the existence of an award can introduce the raw data as evidence to establish the authenticity of a multisig transaction. Thus, like electronic signatures, a court may hold that the raw data of a multisig transaction is compelling proof to establish the existence of an award that has the functional equivalence of an award in writing.

## C. COULD A MULTISIG E-AWARD BE DULY AUTHENTICATED?

Before considering the issue of authentication, it should be pointed out that a multisig e-award is different from a traditional paper award in that it is self-enforcing. Once the multisig signature threshold has been met and the transaction is broadcast to the network, the funds are released to the recipient. Enforcement of a mutlisig award is not necessarily an issue, but whether the act of releasing funds puts an end, in whole, or in part, a dispute and if the resulting raw transaction data could be recognized as an award.

This is important because if multisig is not capable of being recognized, then a multisig transaction could be challenged in any court with competent jurisdiction and an arbitrator's finding reversed. Although technically the multisig transaction is non-reversible at the code level, a court could order compensation to be paid notwithstanding the execution of the transaction on a blockchain. When a participant deposit an asset into a multisig account, there is an understanding that the asset will not be transferred unless certain conditions are met, namely the requirement of M-of-N signatures and the occurrence of an event or the performance of an obligation. The arbitrator makes a determination if the conditions have been satisfied and releases the funds by signing a transaction with his or her private key. As such, multisig can be viewed as a separate and distinct contract from the underlying agreement between the participants. For instance, in a sale of goods agreement, if no goods or sub-standard goods are delivered, the buyer could sue the seller for breach even if a third party releases funds from a multisig account. The decision of the arbitrator, *ipso facto*, does not have *res judicata* effect and is somewhat analogous to a decision of a bank releasing funds under a letter of credit.

---

36 Garner, Bryan A. (Editor) (2009) Black's Law Dictionary, 9. Edition, Eagan, West.
37 Bryant, Randal E./O'Hallaron, David R (2015) Computer Systems: A Programmer's Perspective, 3. Edition, London, Pearson. In C programming language, letters could be written in upper or lower case.

In addition, if the multisig process is not seen as a form of international commercial arbitration and the arbitrator fails to perform the decision-making function to the level expected of a technologically-savvy prudent adjudicator, he or she may face unlimited personal liability. The arbitrator owes concurrent contractual, non-contractual, and equitable duties towards multisig participants. So, it is vital for a multisig transaction to be recognizable as a final and binding award.

Under Article IV(a) the New York Convention, a party applying for recognition must supply a "duly authenticated original award." The Convention does not define the meaning of authentication, but commentators are in general agreement that it can be defined as the process of confirming the authenticity of arbitrators' signatures[38]. The law applicable to issues related to authentication can either be the *lex loci arbitri* or the law of where recognition is sought[39]. According to the *travaux préparatoires,* the drafting committee preferred to give the court where recognition is sought more flexibility in determining the governing law than the Geneva Convention of 1927, which only pointed to *lex loci arbitri*[40]. In common law jurisdictions, the act of confirming the authenticity of a signature is viewed as an evidentiary matter. It does not require to be in any particular form and in the interest of justice and efficient conduct of court business, a simple straightforward method is usually preferred[41]. The act can be performed by public officials, lawyers, witnesses, or arbitration institutions[42]. On the other hand, in most civil law jurisdictions, the concept of authenticity is viewed more narrowly and traditionally require a competent public authority or a notary public to verify the truthfulness of documents[43].

According to *Yu* and *Nasir,* Article IV should be read in conjunction with Article III and if the country of where the award was made permits the issuance of awards in electronic form, there should be no barrier to accepting the award as duly authenticated[44]. *Lederer* notes that although an indefinitely reproducible electronic document cannot be seen as "original," in relation to electronic data, it can be an authentic copy, where the authorship of the data can be reliably proven[45]. However, *Otto* cautions against electronic awards as there is no unanimous view of authentication and some countries have taken a strict view[46]. There is a risk if recognition is sought in a civil law jurisdiction, the multisig transaction may not be viewed as final and binding.

---

38  ICCA (2011), *supra* note 5 at pg. 109.
39  Oberster Gerichtshof, Sep. 3 2008 (O Ltd., et al. v. C Ltd.), in Yearbook Commercial Arbitration XXXIV 409-417 (Albert Jan van den Berg, ed., 2009). However, not all jurisdictions follow this view, some require authentication in conformity with the *lex fori*, such as in Italy. *See* Otto, Dirk 'Article IV': Kronke, Herbert/Nacimiento, Patricia, Otto, Dirk/Port Nicola Christine (Editors) (2010) Recognition and Enforcement of Foreign Arbitral Awards: A Global Commentary on the New York Convention, Alphen aan den Rijn, Kluwer Law International, Wolters Kluwer, pg. 144.
40  E/2704 - Report of the Committee on the Enforcement of International Arbitral Awards (Resolution of the Economic and Social Council establishing the Committee, Composition and Organisation of the Committee, General Considerations, Draft Convention) 14 (1955).
41  Promontoria (Henrico) Ltd v. James Friel, [2019] CSOH 2 (U.K.).
42  Otto, Dirk 'Article IV': Kronke, Herbert/Nacimiento, Patricia, Otto, Dirk/Port Nicola Christine (Editors) (2010) Recognition and Enforcement of Foreign Arbitral Awards: A Global Commentary on the New York Convention, Alphen aan den Rijn, Wolters Kluwer, pg. 183.
43  UNCITRAL 'Promoting Confidence In Electronic Commerce: Legal Issues On International Use Of Electronic Authentication And Signature Methods' (2007) <https://uncitral.un.org/sites/uncitral.un.org/files/media-documents/uncitral/en/08-55698_ebook.pdf> l.a.d. 04/17/2019, pg. 4.
44  Yu, Hong-Lin/Nasir, Motassem (2003) 'Can Online Arbitration Exist Within the Traditional Arbitration Framework?', Journal of International Arbitration, pg. 472.
45  Lederer, Nadine (2019) 'The Enforcement of Cross-Border Online Arbitral Awards and Online Arbitration Agreements. The New York Convention and the Internet: Friends or Foes?' Spain Arbitration Review, pg. 73.
46  Otto, Dirk (2010), *supra* note 42 at pg. 177.

If we apply the law of England as the *lex loci arbitri*, under Section 7(3) of the ECA 2000, "any person" can certify an electronic signature with a statement confirming the authenticity. Section 7, Part III (15)(2)(a), further provides that the term "authentication" relates to whether an electronic communication or data: (i) comes from a particular person or other source; (ii) is accurately timed and dated; (iii) is intended to have legal effect.

In the case of the example in Appendix A, a signature hash can be linked to a particular signer and any alteration to raw transaction data can be detectable. The time and date is part of the data and the intent for the electronic signature to have legal effect can be established by referring to the arbitration agreement and the subsequent conduct of the parties[47]. An expert witness could certify the authenticity of the electronic signatures by checking the embedded metadata. Alternatively, an arbitrator has the option to self-certify as to the authenticity of the transaction. This method of authentication is consistent with the interest of having a pragmatic, flexible, non-formalistic, arbitration-friendly approach to the interpretation of Article IV[48].

## IV. CAN MULTISIG ARBITRATION BE SEEN AS AN AUTONOMOUS LEGAL ORDER?

*Ortolani* has argued that the Bitcoin multisig process is not compatible with any of the traditional narratives of international arbitration, because it offers a level of privacy and enforcement solely depends on the mechanical operation of the protocol without the need for cooperation with national legal systems[49]. Transactions can take place between anonymous users who do not disclose their identity or geographical location[50]. Bitcoin was designed as a reaction to perceived inadequacies of the traditional banking system and is a rejection of a states' exclusive authority over money[51]. Similar to how *lex mercatoria* developed within a community of merchants, Bitcoin users seek self-regulation and stateless mechanisms for resolving disputes[52]. *Ortolani* asserts having the ability to enforce norms is a key indicator of an autonomous legal order, so the multisig process must be regarded as an a distinctive system of dispute resolution[53]. However, this view is not supported by empirical evidence.

Bitcoin provides very little, to no, privacy protection. It is possible to link IP addresses to Bitcoin transactions through network analysis. When a user initiates a Bitcoin client, it sends out a message containing the user's IP address information to other nodes. The software then downloads a list of known peers from other users. This makes it possible to map out IP addresses to Bitcoin addresses[54]. In a study of relay patterns, several hundred high-confidence (>90%) Bitcoin address-to-IP pairings were discovered[55]. In another study, researchers were able to link numerous users to IP addresses, the vast majority of which had probabilities above 0.9, with the average value of pairings at 95.52%[56]. Even with Tor, an anonymizing overlay network, Bitcoin addresses could be linked retroactively with publicly available information. In a case study that crawled thousands of Tor hidden services, social media accounts, and forums, researchers were able to link a number of unique users to Tor hidden services[57]. For some users, they were able to discover personally identifiable information, such as name, gender, age, and location[58]. The researchers concluded that Bitcoin addresses should always be assumed to be compromised as it could be used to de-anonymize users[59]. Moreover, courts have the power to order third party service providers to disclose information about users. Know-your-customer compliant exchanges and custodial wallet providers could be compelled under a pre-action Norwich Pharmacal Order to disclose names, emails, postal addresses, and IP addresses of registered users[60]. In a case from Ireland, a court noted when ordering an internet service provider to disclose information about their subscribers "that whether the right to confidentiality arises by statute or by contract or at common law, it cannot be relied on by a wrongdoer or a person against whom there is evidence of wrongdoing to protect his or her identity… right to privacy or confidentiality of identity must give way where there is *prima facie* evidence of wrongdoing[61]." Thus, with the development of standardized Bitcoin investigative tools that can link transactions to geographical locations and courts' inherent power to order the disclosure of information from service providers, information about Bitcoin users can be discovered and they are not anonymous.

---

47 Trans-Lex.org 'No. IV.5.1 - Intentions of the parties' <https://www.trans-lex.org/924000> l.a.d. 04/17/2019.
48 5A_754/2011, Federal Tribunal, 2 Jul. 2012 (Switz.) (in interpreting Article IV(2) of a partial translation, court applied a flexible, pragmatic and non-formalistic approach).
49 Ortolani, Pietro (2016) 'Self-Enforcing Online Dispute Resolution: Lessons from Bitcoin' Oxford Journal of Legal Studies, I3, V36, pp. 595-629.
50 *Id.* at pg. 612.
51 *Id.*
52 *Id.* at pg. 614.
53 *Id.* at pg. 615.
54 Lr, Andrew/Ao, Douglas (2018) 'Bitcoin Investigations: Evolving Methodologies and Case Studies' Journal of Forensic Research, I3, V9 <https://www.omicsonline.org/open-access/bitcoin-investigations-evolving-methodologies-and-case-studies-2157-7145-1000420-101691.html> l.a.d. 04/17/2019.

Furthermore, Bitcoin users are not purely interested in self-regulation and stateless mechanisms. Users' views range the whole political spectrum. According to a State of Blockchain Report, it was reported that 35% of Bitcoin supporters are Socialist or Liberal, 46% Conservative or Libertarian, while only a small minority, 9%, identified as "Anarcho-Capitalist," a political philosophy that advocates for the elimination of the state[62]. In another survey of blockchain developers, it was found that the majority of respondents indicated they are primarily motivated by money, passion for coding, attraction to the blockchain technology, learning, and community recognition[63]. Bitcoin users have a wide array of views and, in general, do embrace the Westphalian model.

---

55  Koshy, Philip/Koshy, Diana/McDaniel, Patrick 'An Analysis of Anonymity in Bitcoin Using P2P Network Traffic': Safavi-Naini, Reihaneh/Christin, Nicolas (Editors) (2014) Financial Cryptography and Data Security - 18th International Conference, pp. 469-485.
56  Juhász, Péter L./ Stéger, József/Kondor, Dániel/Vattay, Gábor (2018) 'A Bayesian Approach to Identify Bitcoin Users' PLoS ONE, I:12, V:13 <https://doi.org/10.1371/journal.pone.0207000>.
57  Al Jawaheri, Husam/Al Sabah, Mashael/Boshmaf, Yazan/Erbad, Aiman (2018) 'When A Small Leak Sinks A Great Ship: Deanonymizing Tor Hidden Service Users Through Bitcoin Transactions Analysis' ArXiv180107501 <http://arxiv.org/abs/1801.07501> l.a.d. 04/17/2019.
58  *Id.* at pg. 2.
59  *Id.* at pg. 10.
60  Yalabık, Fulya Teomete/Yalabık, İsmet (2019) 'Anonymous Bitcoin v Enforcement Law' International Review of Law, Computers & Technology, I:1, V:3, DOI: 10.1080/13600869.2019.1565105, pg. 48.
61  EMI Records Ltd., Sony BMG Music Entertainment Ltd., Universal Music Ltd., Warner Music Ireland Ltd v. Eirecom Ltd., BT Telecommunications Ltd., [2006] E.C.D.R. 5 (Ire.). The test is where a person is wittingly or unwittingly caught up in the wrongdoing of another so as to facilitate that wrongdoing and is more than a mere witness, a court may order the person to provide the injured party the necessary information to enable the ultimate wrongdoer to be identified. The wrongful act could arise out a crime, tort, breach of contract, equitable wrong, or contempt of court and there must be a need for an order to enable action in the interest of justice. *See* ArcelorMittal USA LLC v. Essar Steel Ltd., [2019].
62  Bauerle, Nolan/Ryan, Peter (2018) 'CoinDesk Releases Q2 2018 State of Blockchain Report CoinDesk' <https://www.coindesk.com/state-of-blockchain-q2-2018> l.a.d. 04/17/2019.

And lastly, Bitcoin multisig cannot be viewed as a customary norm in *lex mercatoria*. To be recognized as a norm, the practice must persist over a substantial period of time, reflect trade habits and market usages, be universal, be extrinsic to the legal system, and be of utilitarian benefit for a merchant community[64]. The BIP-11 M-of-N Standard Transactions proposal was accepted December 13, 2011 and BIP-16 Pay to Script Hash, which enabled multisig scripting, was accepted April 1, 2012[65]. The first multisig wallet service was launched in August 2013[66]. According to statistics of 2/3 multisig account usage, the number of deposited funds picked up in late 2015, but has since leveled off after the 2017 Bitcoin "tulip bubble," *see* Appendix B. As it can be seen multisig is a relatively recent technology that does not have wide usage over a substantial period of time, so customs within the community cannot be elevated to the level of *lex mercatoria*. Furthermore, there is no evidence to show that Bitcoin multisig is currently being used by a concrete and identifiable merchant community.

Given these points, multisig ought not be seen as a distinctive legal order. It would be more advantageous to bring the multisig process under the purview of international commercial arbitration law. International arbitration has a wealth of rules and case law for arbitrators to draw from. Awards can be recognized by numerous national legal systems under the New York Convention[67] and arbitrators, acting in a quasi-judicial capacity, are afforded immunity from liability under national laws and arbitration rules[68].

## V. CONCLUSION

Multisig is an exciting technology that could change how merchants resolve cross-border disputes. The multisig protocol functions similar to escrow, but a key difference is that no one party, including the arbitrator, has sole possession of the funds. In a 2/3 multisig account, at least two parties must authorize a transfer by signing data with their respective private keys. This scheme provides a more efficient and trust-less mechanism for trading. However, there is some uncertainty whether or not multisig can be recognized as an arbitral award in some jurisdictions. The New York Convention requires awards to be "duly authenticated." In civil law systems, the authentication of signatures is traditionally performed by public authorities or notaries. Moreover, the *lex arbitri* may impose additional form and document legalization requirements.

If parties select the law of England as the seat of arbitration, they are free to agree on the form of an award under Section 52(1) of the Arbitration Act 1996. Through the concept of party autonomy, parties can restrict arbitrators from following guidelines on the content of awards and authorize the issuance of an award in electronic form. The parties should explicitly opt-out of a reasoned award as provided for under some arbitration rules. In addition, names and wallet addresses should be associated with parties at the time when the arbitration agreement is made. By doing this, an expert witness is able to certify the identity of parties and authenticate electronic signatures. Under ECA 2000, Section 7(1)(a), electronic signatures are admissible as evidence and under section 7(3), any person can certify the authenticity of such signatures.

---

[63] Bosu, Amiangshu/Iqbal, Anindya/Shahriyar, Rifat/Chakroborty, Partha (2018) 'Understanding the Motivations, Challenges and Needs of Blockchain Software Developers: A Survey' ArXiv181104169 <http://arxiv.org/abs/1811.04169> l.a.d. 04/17/2019.
[64] Bhala, Raj (1996) 'Applying Equilibrium Theory and the FICAS Model: A Case Study of Capital Adequacy and Currency Trading' Saint Louis University Law Journal, V41, I:125, pp. 205-206.
[65] Bitcoin Project 'Bitcoin Improvement Proposals' <https://github.com/bitcoin/bips> l.a.d. 04/17/2019.
[66] O'Brien, Will (2014) 'How 2014 Became the Year of Multisig' CoinDesk <https://www.coindesk.com/2014-became-year-multisig> l.a.d. 04/17/2019.
[67] As of 02 April 2019, there are 159 contracting states of the New York Convention.
[68] U.K. Arbitration Act, 1996, Section 29(1); ICC Rules of Arbitration (2017), Article 41; Trans-Lex.org 'Principle XIII.2.7 - Immunity of arbitrator' <https://www.trans-lex.org/970032> l.a.d. 04/17/2019.

Multisig ought to be viewed as compatible within the international arbitration legal order, where the process is defined in an arbitration agreement and the raw transaction data is the actual arbitral award in an electronic form. Much in the same way courts have been able to adapt the law regarding paper contracts to the emergence of technologies like telegram, telex, fax, and e-mail, blockchain-based applications will be another method for entering into contractual relations. The technological novelty of multisig does not justify characterizing it as a separate legal order.

Although multisig is compatible with the aims of international arbitration, in that it could provide a fast, cost-effective, and final method for resolving disputes privately, it is unlikely it will be widely used at the present. For one, there is a steep learning curve for merchants and their legal representatives to learn how to adequately use the technology. Secondly, arbitration institutions might be reluctant or slow to modernize rules to allow for conducting arbitration by electronic means. And finally, the multisig protocol is not intuitive enough for ordinary commercial use.

# Appendix A: Multisig Transaction on Bitcoin Testnet Example

The author selected Bitcoin, a popular virtual currency, to illustrate a multisig transaction. Bitcoin Core version 0.17.1 was used. For simplicity, we selected an already completed testnet multisig transaction and looked-up the raw transaction ("TxRaw") data from a blockchain explorer[69]. We then decoded raw data with the Bitcoin client, filtering out the relevant input information as follows:

```
./bitcoin-cli -testnet -named decoderawtransaction hexstring=TxRaw | jq '.vin | map(.scriptSig.asm | split(" ")[3])'
[ "52210279aac3e06ee2e54ab5952a75fe742883d5ecaa2da33dfeb60a6940a435ed53992102f90212cad84ab0875ef34d17c09e5f
ecff34f25786f99ddb5f4bdca5c599707b2103d3009499b501c7be0f4f7d3d8f45af4d2dd9104070e7dfedb5e57949a10a09af53ae"
]

./bitcoin-cli --testnet -named decodescript
hexstring=52210279aac3e06ee2e54ab5952a75fe742883d5ecaa2da33dfeb60a6940a435ed53992102f90212cad84ab0875ef34d1
7c09e5fecff34f25786f99ddb5f4bdca5c599707b2103d3009499b501c7be0f4f7d3d8f45af4d2dd9104070e7dfedb5e57949a10a09
af53ae | jq '.reqSigs,.type, .addresses'
2 "multisig" ["mzV1dsMdDjtLSfRa2rPrE2oJpRtynKkjJX", "mpGZniUmoCemQzRbazvdgzGkmjUQ3fZN8L",
"n2dSPmt5cv2hFNfQqoZtvRJ6bZmypNBSvH"]
```

The above information (emphasized in bold) tells us that the transaction utilized the multisig protocol and required the signature of two private keys out of three listed wallets. Next, we used the same raw data to extract relevant output information:

```
./bitcoin-cli -testnet -named decoderawtransaction hexstring= TxRaw | jq '.vout | map(.scriptPubKey.asm)'
["OP_RETURN
412d4a6f686e536d6974682d4b6b6a4a5820432d41636d652d665a4e384c20522d42616b65722d4e42537648204c6f6e646f6e2
0436661376a6168445444566a5a774b55706b3777317970786738733d"]
echo
"412d4a6f686e536d6974682d4b6b6a4a5820432d41636d652d665a4e384c20522d42616b65722d4e42537648204c6f6e646f6e2
0436661376a6168445444566a5a774b55706b3777317970786738733d" | xxd -ps -r; echo
A-JohnSmith-KkjJX C-Acme-fZN8L R-Baker-NBSvH London Cfa7jahDTDVjZwKUpk7w1ypxg8s=
```

This metadata (emphasized in bold) helps us associate the previously extracted wallet addresses from the input information to a name. The alphanumeric string after London is the last 28 characters of a digital signature that signed the message. The whole signature hash can be reproduced by signing the text with the wallet that ends with KkjJX (John Smith's wallet). The validity of the signature can be confirmed with the Bitcoin client by plugging in the signer's wallet address, signature hash, and message.

```
./bitcoin-cli -testnet signmessage "mzV1dsMdDjtLSfRa2rPrE2oJpRtynKkjJX" "A-JohnSmith-KkjJX C-Acme-fZN8L R-
Baker-NBSvH London"
IO0vDf3ZqfRZ8FGGsnzzkMc65YQWIWb2+YqcQ9j/APK2QN1E2TTV/3xPkThhCfa7jahDTDVjZwKUpk7w1ypxg8s=

./bitcoin-cli -testnet verifymessage "mzV1dsMdDjtLSfRa2rPrE2oJpRtynKkjJX"
"IO0vDf3ZqfRZ8FGGsnzzkMc65YQWIWb2+YqcQ9j/APK2QN1E2TTV/3xPkThhCfa7jahDTDVjZwKUpk7w1ypxg8s="
"A-JohnSmith-KkjJX C-Acme-fZN8L R-Baker-NBSvH London"
true
```

---

69  Bitaps.com (2019) 'Bitcoin Testnet Transaction'
    <https://tbtc.bitaps.com/fa65bc5fa0ee39e012282701a4ce378474183a330487e839cd1b65b398d7646e> l.a.d.
    04/17/2019.

With the metadata extracted from the raw transaction data, the digital signature provided by one of the parties, and the publicly available information on a blockchain explorer, we can establish with high confidence the following:

- Transaction id *fa65bc5fa0ee39e012282701a4ce378474183a330487e839cd1b65b398d7646e* was completed on 28 March 2019 at 15:46:53 UTC
- The transaction amount was 0.005 BTC
- "A-JohnSmith-KkjJX" relates to *mzV1dsMdDjtLSfRa2rPrE2oJpRtynKkjJX*
- "C-Acme-fZN8L" relates to *mpGZniUmoCemQzRbazvdgzGkmjUQ3fZN8L*
- "R-Baker-NBSvH" relates to *n2dSPmt5cv2hFNfQqoZtvRJ6bZmypNBSvH*
- The transaction makes reference to London.
- Mr. John Smith's wallet digitally signed the embedded data.
- The record is unaltered given the number of confirmations, which can be seen on any blockchain explorer.

# Appendix B: Breakdown of BTC Stored in 2/3 Multisig Accounts[70]

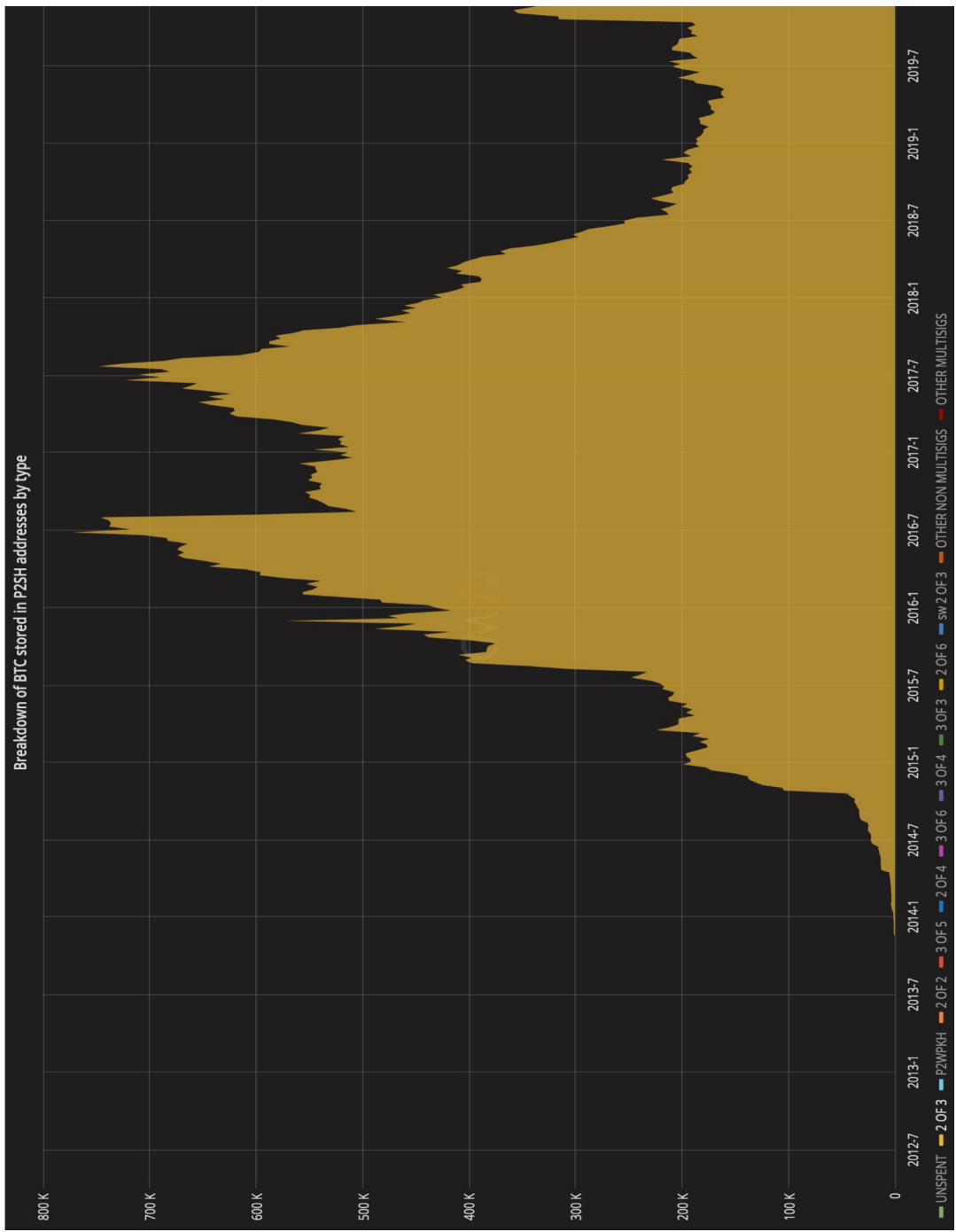

---

70  p2sh.info (2019) 'P2SH repartition by type' <https://txstats.com/dashboard/db/p2sh-repartition-by-type> l.a.d. 11/24/2019.